\documentclass[epj]{svjour}
\usepackage{achemso}
\usepackage{epsfig}
\usepackage{graphicx}
\usepackage{graphics}
\usepackage{indentfirst}
\usepackage{multirow}
\usepackage{latexsym}
\begin{document}

\title{From Layers to Nanotubes: Transition Metal Disulfides TMS$_2$}
\author{Nourdine Zibouche\inst{1} \and Agnieszka Kuc\inst{1} \and Thomas Heine\inst{1}}
\institute{\inst{1}School of Engineering and Science, Jacobs University Bremen, \\
Campus Ring 1, 28759 Bremen, Germany\\
Email: a.kuc@jacobs-university.de}
\date{Received: date / Revised version: date}

\abstract{
MoS$_2$ and WS$_2$ layered transition-metal dichalcogenides are indirect band gap semiconductors in their bulk forms.
Thinned to a monolayer, they undergo a transition and become direct band gap materials.
Layered structures of that kind can be folded to form nanotubes.
We present here the electronic structure comparison between bulk, monolayered and tubular forms of transition metal disulfides using first-principle calculations.
Our results show that armchair nanotubes remain indirect gap semiconductors, similar to the bulk system, while the zigzag nanotubes, like a monolayer, are direct gap materials, what suggests interesting potential applications in optoelectronics.
\PACS{
      {PACS-key}{discribing text of that key}   \and
      {PACS-key}{discribing text of that key}
     }
}

\maketitle

\section{Introduction}
\label{intro}

The discovery of carbon nanotubes\cite{Iijima1991} immediately stimulated the research to explore other inorganic materials, which may form tubular structures.
Nanoparticles of various inorganic layered materials, such as WS$_2$ or MoS$_2$, in analogy to carbon, can form nanostructures of fullerene-like and nanotubular shapes.
In 1992 and 1993, Tenne and co-workers have shown that layered transition-metal dichalcogenides (LTMDCs), namely WS$_2$ and MoS$_2$, form the so-called inorganic nanotubes and fullerene-like nano\-particles.\cite{Tenne1992, Margulis1993}
Nowadays, it is well-known that several layered inorganic compounds posses structures comparable to gra\-phite (honey-comb-like) and as such can form tubes.
Among them are transition metal dichalcogenides, halides, and oxides.

\begin{figure*}[t!]
\begin{center}
\includegraphics[scale=0.40,clip]{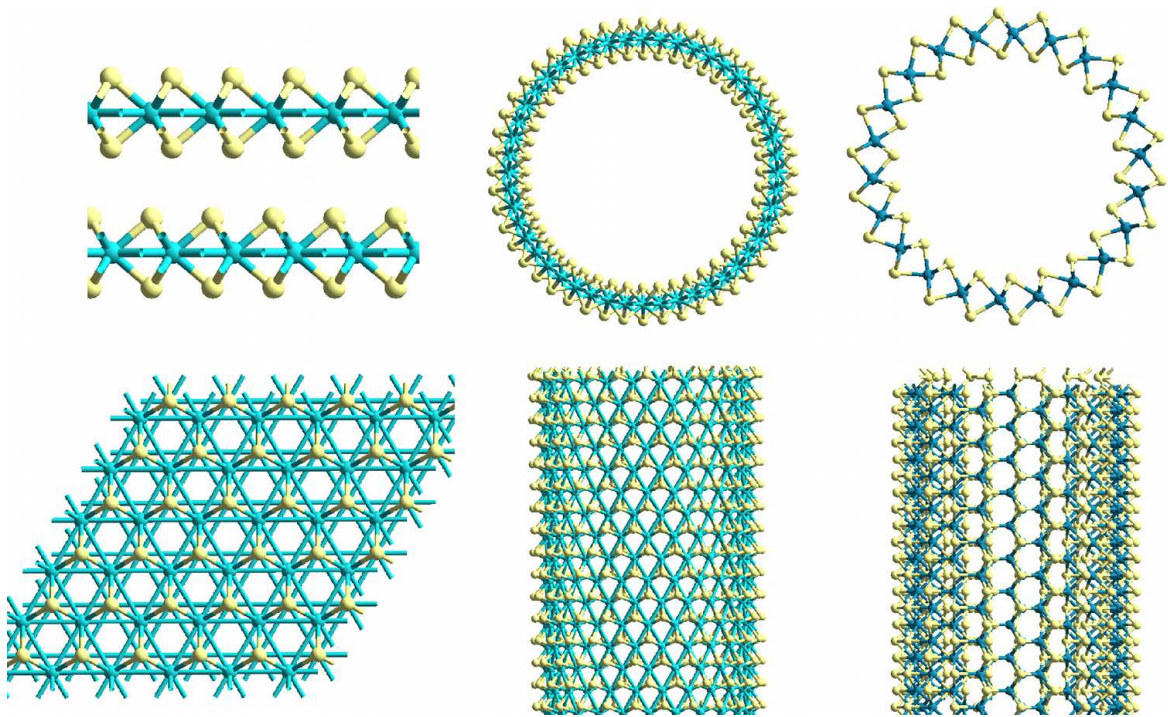}
\caption{\label{fig:1}(Color online) The atomic structures of layered (left) and tubular (middle -- zigzag and right -- armchair) transition-metal disulfides of TMS$_2$ type (TM = Mo, W). (top) the cross sections and (bottom) the hexagonal sheets are shown for both types of the systems.}
\end{center}
\end{figure*}

LTMDCs of TMX$_2$ type (TM = Mo, W, Nb, Re, Ti, etc., X = S, Se, Te) have been extensively studied experimentally and theoretically for the last 40 years.\cite{Wilson1969, Mattheis1973, Kam1982, Coehoorn1987, Kobayashi1995, Wilcoxon1997, Reshak2005, Lebegue2009, Arora2009, Splendiani2010, Mak2010, Kuc2011}
Molybdenum disulfide is a prototypical LTMDC, which is composed of two-dimensional trilayered S-Mo-S sheets stacked on top of one another and held together by weak bonds (see Fig.~\ref{fig:1} left).
Each sheet is trilayered with a Mo atom in the middle that is covalently bonded to six S atoms located in the top and bottom of the sheet (see Fig.~\ref{fig:1}).

LTMDCs are of special interest because of their extraordinary properties and diverse applications.
We have recently shown that TMS$_2$ (TM = Mo, W) are indirect band gap materials in their bulk form and they become direct band gap semiconductors when thinned to monolayers.\cite{Kuc2011}
These results support the recent experimental findings for MoS$_2$\cite{Splendiani2010} and show that quantum confinement is the key factor to tune electronic structures of LTMDCs.
Therefore, electronic and optical properties of LTMDCs are very interesting for fabrication of field-effect transistors, as it was recently realized by Kis and co-workers for a MoS$_2$ monolayer.\cite{Radisavljevic2011, Bertolazzi2011}
In this transistor, HfO$_2$ was used as a gate insulator.\cite{Radisavljevic2011}
It exhibits a room-temperature current on/off ratio exceeding 1x10$^8$ and mobility comparable to the mobility of thin silicon films or graphene nanoribbons.
Other applications, such as catalysis, optoelectronics and photovoltaics, have been proposed and investigated as well.\cite{Kam1982, Tenne1985, Coehoorn1987, Sienicki1996, Gourmelon1997}

Bulk LTMDCs can be synthesized using a variety of methods.
Generally, dichalcogenides are prepared by heating under hydrogen sulfide flow.
Monolayers of such layered materials can be produced by e.g.\ liquid exfoliation, as successfully performed by Coleman et al.\cite{Coleman2011}
Moreover, it was shown that it is possible to distinguish the number of layers of dichalcogenide crystals, using a simple optical model, by calculating the contrast of nanolayers deposited on SiO$_2$ wafer with varying thickness.\cite{Benameur2011}

Transition metal dichalcogenide nanotubes (TMDC-NTs) can be produced using e.g.\ chemical vapor transport technique\cite{Remskar1998} or by high-temperature annealing of the respective metal trisulfides.\cite{Nath2001}
For more details, refer to the comprehensive review on inorganic nanotubes by Nath et al.\cite{Rao2003}

TMDC-NTs have been a subject of various investigations\cite{Tenne1992, Seifert2000, Milosevic2007, Tenne2010} in order to understand physical and chemical properties of these materials and to propose important applications.
For example, it has been shown that TMDCs-NTs behave as exceptional lubricants.\cite{Mattheis1973a, Drummond2001}
The mechanical properties of WS$_2$ nanotubes under axial tension and compression,\cite{Kaplan2006} and MoS$_2$ nanotubes under squeezing,\cite{Stefanov2008} have been investigated in terms of the lubrication process at high loads.
Experimental results show that at high loads the lubricant is expelled from the location of close interface distance due to mechanical pressure, and it hence looses its tribological properties, leading to an increased friction coefficient.
If the MoS$_2$ nanotubes or -onions are added to base grease, the friction coefficient remains low, even at very high loads.
Density functional--based calculations show that under the squeezing of nanotubes the MoS$_{\rm 2}$ platelets are formed, partially attached to the grips, which provide good lubrication at the position of closest contact of the materials sliding against each other.
This excellent lubrication of nanostructures is interpreted as 'nano-coating'.
The mechanical behaviour of WS$_{\rm 2}$ nanotubes under axial tension and compression\cite{Kaplan2006} shows that they are ultra-strong and elastic, what distinguishes them from other known materials.
The failure of the nanotubes is abrupt starting at a single atomic defect and propagating very quickly across its entire circumference.
Moreover, MoS$_2$-NTs have been recently used for catalytic conversion of carbon oxide and hydrogen into methane and water.
These foundings are quite unexpected, as the fully bonded sulfur atoms in the LTMDC surfaces are not expected to be chemically active.

Doping these semiconducting nanotubes may lead to new optoelectronic nanomaterials.
Ivanovskaya et al.\cite{Ivanovskaya2006a} have investigated the effect of Mo to Nb substitution on the electronic structure of MoS$_2$ nanotubes using density functional based tight-binding (DFTB) method.\cite{Seifert1996, Oliveira2009}
It has been found that composite Mo$_{1-x}$Nb$_x$S$_2$ nanotubes (with Nb contents of 5, 10 and 25 at\%) are more stable than the corresponding pure tubes.
This effect was even stronger for larger tube diameters. 
At room temperature the Nb dopant prefers to arrange in ordered manner along the tube axis, whereas the site entropy contribution favors a random distribution at high temperatures.
Ivanovskaya et al.\cite{Ivanovskaya2006a} reportred that all the doped nanotubes studied in this work have metallic properties, independent of their chirality, diameters or the substitutional patterns.
The density of states (DOS) close to the Fermi level of Nb-substituted MoS$_2$ nanotubes can be tuned in a wide range by the degree of doping.
On the other hand, Ivanovskaya et al.\cite{Ivanovskaya2008} have also studied the Nb substitution in MoS$_2$ layered materials and found out that the Nb atoms prefer to distribute homogeneously.

The properties of inorganic nanotubes can be further widely explored, as they are much less investigated as their carbon counterparts.
Especially, understanding of the change in electronic properties, when going from layers to nanotubes or after substitution, is of high interest due to potential applications in nano- and optoelectronics.

In this paper, we would like to present the comparison of the electronic structure between bulk, monolayered and tubular forms of TMS$_2$, where TM = Mo and W (see Fig.~\ref{fig:1} middle and right).
The first principle calculations were performed using localized Gaussian basis functions and compared to the available experimental and theoretical data.
Our results show that TMDC-NTs of armchair type are indirect band gap semiconductors (like a bulk structure), while their zigzag counterparts have direct band gaps and resemble the electronic structure of a TMDC monolayer.

\section{Methods}
\label{Sec:Methods}

In this work, we have studied TMDC-NTs of TMS$_2$ type (where TM = Mo and W) with different tube sizes for both armchair and zigzag configurations.
All layered structures, initial systems to built up nanotubes, have hexagonal symmetry and belong to the $P6_3/mmc$ space group.
The monolayers were cut out from the fully optimized bulk structures as (0 0 1) surfaces.

First-principle calculations were performed on the basis of density functional theory (DFT) as implemented in the CRYSTAL09 code.\cite{CRYSTAL09}
The exchange and correlation terms were described using general gradient approximation (GGA) in the scheme of PBE (Perdew-Burke-Ernzerhof).\cite{PBE}
The following basis sets were used: Mo$\_$SC$\_$\-HAY\-WSC-311(d31)G$\_$cora$\_$1997 (for Mo atoms),\cite{Cora1997} W$\_$co\-ra\-$\_$\-1996 (for W atoms),\cite{Cora1996} and S$\_$86-311G*$\_$lichanot$\_$1993 (for S atoms).\cite{Lichanot1993}
We have already shown, that the combination of these bases and PBE functional gives very good results for electronic structure calculations of layered TMDCs.\cite{Kuc2011}

The shrinking factor for bulk and layered structures was set to 8, what results in the corresponding number of 50 and 30 $k$-points in the irreducible Brillouin zone, respectively.
The tubular structures were treated with the shrinking factor 8 and 4, that correspond to 5 and 3 $k$-points along the tube axis, respectively.
The mesh of $k$-points was obtained according to the scheme proposed by Monkhorst and Pack.\cite{Monkhorst1976}
Band structures were calculated along the high symmetry points using the $\it{\Gamma-M-K-\Gamma}$ and $\it{\Gamma-K-\Gamma}$ paths for bulk/monolayer and nanotubes, respectively.

The bulk, layers and nanotube calculations have been carried out employing 3D, 2D and 1D periodic boundary conditions, respectively.

Optimization of initial experimental structures was performed using analytical energy gradients with respect to atomic coordinates and unit cell parameters within a quasi-Newton scheme combined with the BFGS (Broyden-Flet\-cher-Goldfarb-Shanno) scheme for Hessian updating.
The optimized lattice parameters for all the studied materials are given in Table~\ref{tab:1}.
\begin{table}
\caption{\label{tab:1}Calculated and experimental lattice parameters [\AA] of 2D hexagonal transition-metal dichalcogenides in the form of TMS$_2$ (TM = Mo, W). Results obtained at the DFT/PBE level. }
\begin{tabular}{ccccc}
\hline\noalign{\smallskip}
\multirow{2}{*}{\textbf{~Structure~}} & \multicolumn{2}{c|}{\textbf{Theory}} & \multicolumn{2}{c}{\textbf{Exp.\cite{Wilson1969, Mattheis1973, Coehoorn1987a}}} \\
\cline{2-5}
                   & \textbf{$a$} & \textbf{$c$} & \textbf{$a$} & \textbf{$c$} \\
\noalign{\smallskip}\hline\noalign{\smallskip}
\textbf{MoS$_2$} &~~3.173~~&~~12.696~~&~~3.160~~&~~12.295~~\\
\textbf{WS$_2$}  &~~3.164~~&~~12.473~~&~~3.154~~&~~12.362~~\\
\noalign{\smallskip}\hline
\end{tabular}
\end{table}

\section{Results and Discussion}
\label{Sec:ResDis}
\subsection{Structural Properties}
\label{Sec:Strut}

Our calculations were performed for prototypical (n,0) zigzag and (n,n) armchair TMS$_2$ nanotubes (TM = Mo and W) as function of n.
The tube diameters ($d$) range between 12.4~\AA\ and 24.9~\AA\ for zigzag NTs, and 19.2~\AA\ and 42.3~\AA\ for armchair ones.
These correspond to the index n = 11--24.

\begin{figure}[h!]
\begin{center}
\includegraphics[scale=0.65,clip]{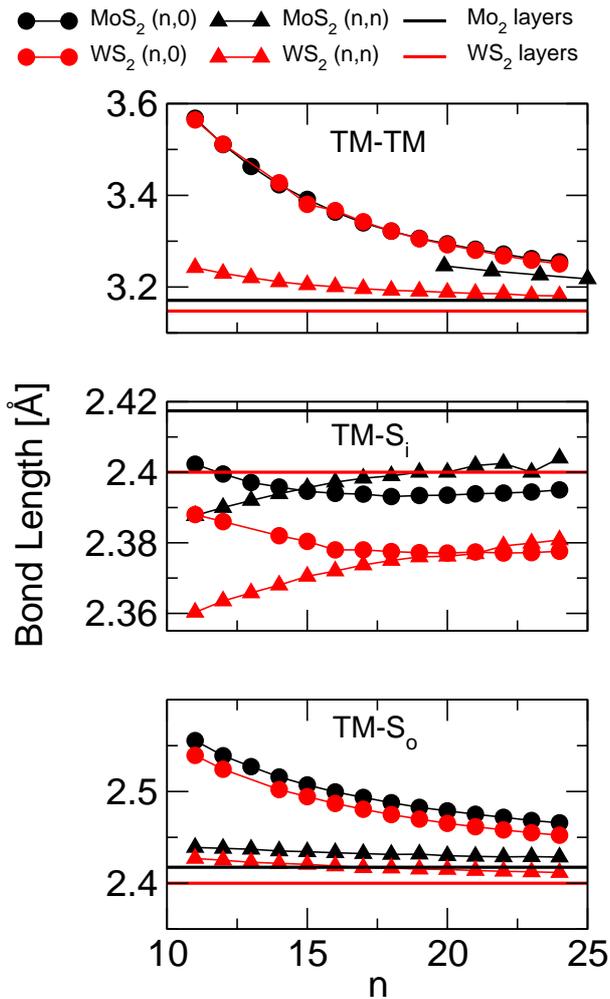}
\caption{\label{fig:2a}(Color online) Selected bond lengths of zigzag and armchair TMS$_2$-NTs (TM = Mo, W) as function of the chiral index n.}
\end{center}
\end{figure}

We have compared the structural properties of different NTs with those of bulk/monolayered systems (see Fig.~\ref{fig:2a}).
The lattice parameter $a$ changes for zigzag forms and we found an increase by around 0.14~\AA\ when increasing $d$. 
Generally, we find the Mo-S bonds longer that their W-S analogues (up to 0.1~\AA\ depending on the bond type).
The zigzag topologies provide longer bonds (up to 0.3~\AA\ depending on the tube diameter) than the armchair ones.
Layered structures have one type of TM--S bond lengths, while in the tubular form we can distinguish the inner and outer TM--S bonds (hereafter, we will refer to them as TM--S$_i$ and TM--S$_o$, respectively, unless otherwise stated).

In detail:
The optimized bond lengths of tubes are slightly different than those of the planar sheets.
Rolling up a monolayer into a nanotube causes elongation of the TM--S$_o$ bond lengths by around 0.05--0.14~\AA\ for zigzag and 0.01--0.02~\AA\ for armchair types (see Tab.~\ref{tab:2a}).
On the other hand, the TM--S$_i$ bonds shrink by 0.01~\AA\ for both zigzag and armchair forms.
TM--S$_o$ bonds are longer than TM--S$_i$ by 0.02--0.07~\AA\ for armchair NTs and by 0.07--0.15~\AA\ for zigzag tubes for both disulfides (depending on the tube diameter).
Comparing various tube sizes, these bond lengths behave differently.
Increasing the tube diameter, the TM--S$_i$ bond lengths decrease or increase for armchair and zigzag forms, respectively.
The TM--S$_o$ bond lengths in zigzag TMS$_2$-NTs change for different chiralities, namely they decrease by around 0.09~\AA\ with increasing the tube diameter from (11,0) to (24,0) NT, while they stay almost unchanged in armchair structures.
\begin{table}
\caption{\label{tab:2a}Calculated TM--S bond lengths [\AA] of nanotubular and 2D hexagonal TMS$_2$ (TM = Mo, W). Only the largest and the smallest diameter tubes are considered. For more details see text and Fig.~\ref{fig:2a}.}
\begin{tabular}{cccccc}
\hline\noalign{\smallskip}
\textbf{Bond} & \textbf{(11,0)} & \textbf{(24,0)} & \textbf{(11,11)} & \textbf{(24,24)} & \textbf{2D}\\
\noalign{\smallskip}\hline\noalign{\smallskip}
\textbf{Mo--S$_o$} &~~2.556~~&~~2.466~~&~~2.439~~&~~2.429~~&~~2.417~~\\
\textbf{Mo--S$_i$}  &~~2.402~~&~~2.395~~&~~2.388~~&~~2.404~~&~~2.417~~\\
\textbf{W--S$_o$}   &~~2.539~~&~~2.452~~&~~2.427~~&~~2.412~~&~~2.400~~\\
\textbf{W--S$_i$}    &~~2.388~~&~~2.378~~&~~2.360~~&~~2.381~~&~~2.400~~\\
\noalign{\smallskip}\hline
\end{tabular}
\end{table}

The TM--TM distances (within the tube circumference) are generally longer for tubular forms and decrease with increasing the diameter of the NTs.
Going from (11,0) to (24,0) NT, these values decrease by around 0.06~\AA\ in the armchair configuration and by around 0.30~\AA\ for the zigzag.
Interestingly, the TM--TM distances (in armchair NTs) along the axial direction do not change with the tube diameter and their values are similar to those in the layered systems.
For small tube diameters, the S--S bonds are shorter than the corresponding bonds in the layered structures, but they increase with increasing $d$, almost approaching the values for a bulk and a monolayer.
The same results are found for both disulfides.

\subsection{Energetic Properties}
\label{Sec:Energ}

\begin{figure}[h]
\begin{center}
\includegraphics[scale=0.35,clip]{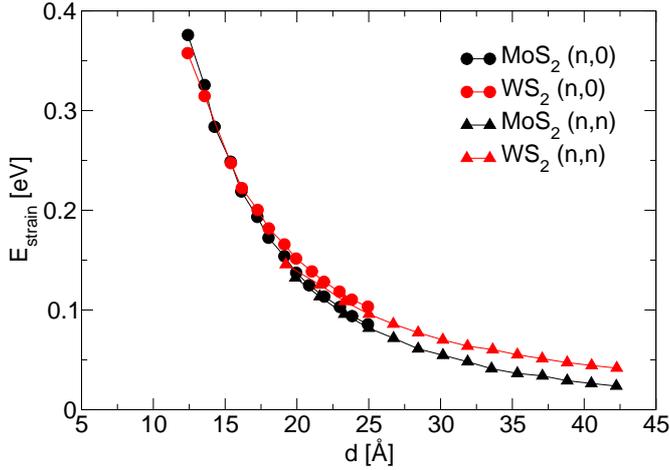}
\caption{\label{fig:2}(Color online) Calculated strain energies of zigzag and armchair TMS$_2$-NTs (TM = Mo, W) as function of the tube diameter ($d$). The energies are given per atom and related to the infinite monolayers of the corresponding disulfides.}
\end{center}
\end{figure}

The calculated strain energies, i.e.\ the differences between the total energies (per atom) of the tubes and the monolayers, scale as 1/$d^2$, where $d$ is the tube diameter (see Fig.~\ref{fig:2} and Tab.~\ref{tab:2}).
The correlation coefficients of the $y=C/d^2$ curves, where C is a constant, are larger than 0.998 for all the studied tubes.
The strain energy scaling is similar to carbon nanotubes (CNTs)\cite{Hernandez1998}, however, the energy values are around 1 order of magnitude larger than for CNTs with similar diameters.
We can understand this result easily, as it is much easier to fold a monoatomic layer (CNTs)  than a triatomic (TMDCs). 
\begin{table}
\caption{\label{tab:2}Correlation coefficients (r$^2$) of the strain energy fitting as function of $C/d^2$ for all studied tubular systems. Constant C is given in [eV \AA$^2$] and d denotes tube diameter [\AA].}
\begin{tabular}{ccc}
\hline\noalign{\smallskip}
\textbf{NT} & \textbf{C} & \textbf{r$^2$} \\
\noalign{\smallskip}\hline\noalign{\smallskip}
\textbf{Zigzag MoS$_2$} &~~57.504~~&~~0.9993~~\\
\textbf{Zigzag WS$_2$}  &~~58.142~~&~~0.9990~~\\
\textbf{Armchair MoS$_2$} &~~50.897~~&~~0.9997~~\\
\textbf{Armchair WS$_2$}  &~~59.681~~&~~0.9985~~\\
\noalign{\smallskip}\hline
\end{tabular}
\end{table}

The strain energies of armchair nanotubes are just slightly more favorable than the zigzag for a given diameter for both disulfides.
This is in a good agreement with the work of Seifert et al.,\cite{Seifert2000} where the authors have used the DFTB method.
The strain energy factor $C$ (in eV \AA$^2$) for armchair and zigzag MoS$_2$ NTs on the DFTB level was 57.613 and 61.5206, respectively (cf. Tab.~\ref{tab:2}).
Also, the molybdenum forms are more stable than the tungsten for diameters larger than 15~\AA\ and this difference grows with the size of the tube.
 
\subsection{Electronic Properties}
\label{Sec:Elect}

\begin{figure}[t!]
\begin{center}
\includegraphics[scale=0.65,clip]{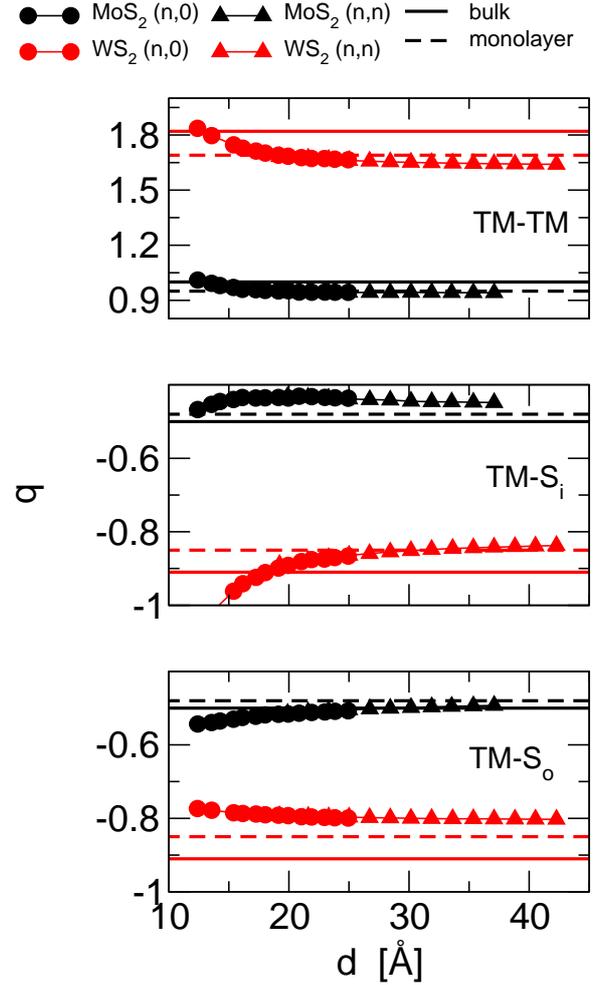}
\caption{\label{fig:3}(Color online) Calculated Mulliken atomic charges ($q$) of tubular and layered forms of TMS$_2$ (TM = Mo and W) as a function of tube diameter ($d$).}
\end{center}
\end{figure}

We have studied the Mulliken charges ($q$) and the electronic band structures of TMS$_2$ (TM = Mo and W) in both layered and tubular forms.
The charge transfer from Mo atoms to S atoms within a monolayer ($q_{Mo}=0.95$, $q_{S}=-0.48$) is similar to that in the bulk form of MoS$_2$ ($q_{Mo}=0.99$, $q_{S}=-0.50$; see Fig.~\ref{fig:3}).
For MoS$_2$, Seifert et al.\cite{Seifert2000} have found $q_{Mo}=0.90$ and $q_{S}=-0.44$.
Slight differences are found for the WS$_2$ layers, where for the TM atom we have the charge of $1.81$ and $1.69$ for the bulk and monolayer, respectively.
The charge of sulfur atoms, $-0.90$ in bulk and $-0.85$ in the monolayer, is larger than in MoS$_2$.

Folding a monolayer of TMS$_2$ into a nanotube affects the atomic charges mostly for small diameter sizes.
Generally, the zigzag and armchair forms have the same values of $q$ for a given tube diameter.
The atomic charges approach a constant value (independent of $d$) for diameters larger than 25~\AA.
Moreover, the tungsten nanotubes deviate from the layered structures much more than the molybdenum ones.

In detail, in TMS$_2$-NTs we should distinguish two different types of sulfur atoms, S$_i$ and S$_o$, that are forming the inner and outer walls of the nanotubes.
In case of MoS$_2$-NTs, the S$_o$ atoms are slightly more negative than the S$_i$ atoms for both zigzag and armchair forms.
Also, S$_o$ atoms approach the atomic charges of layered structures for larger diameters, while in the case of S$_i$ atoms, the $q$ values are close to the layered structures for smallest tube diameters.
The situation changes for WS$_2$-NTs, where the S$_o$ atoms are less negative than the inner ones and the sulfur atoms of the layered forms.
The inner sulfurs, for WS$_2$-NTs $d$ smaller than 15~\AA, have charge close to the one of a bulk structure, while for $d$ larger than 25~\AA\ this values are close to the charges of a monolayer.

The charge of TM atoms becomes less positive when the tube diameter increases.
For the smallest tube sizes, these values are closer to the bulk form and for $d$ larger than 15~\AA, they approach the charges of the monolayers.

\begin{figure}[ht!]
\begin{center}
\includegraphics[scale=0.33,clip]{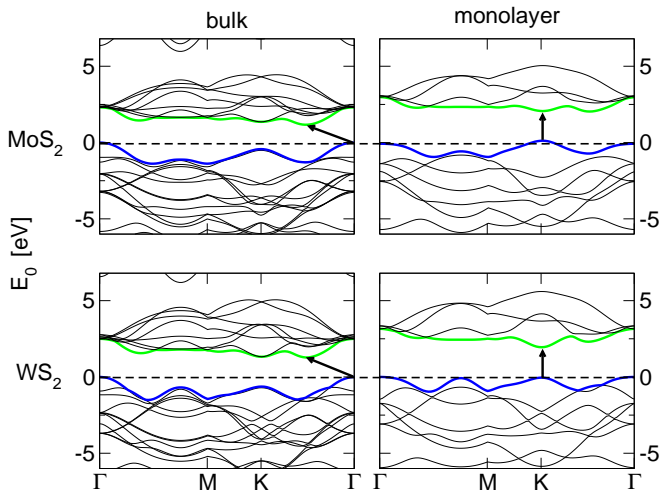}
\caption{\label{fig:4}(Color online) Band structures of bulk and monolayered TMS$_2$ (TM = Mo and W). The arrows indicate the fundamental band gap (direct or indirect; $\Delta$) for a given system. The top of valence band (blue) and bottom of conduction band (green) are highlighted (Online color).}
\end{center}
\end{figure}

Recently, we have studied the electronic structure of layered TMS$_2$ (with TM = Mo, W, Nb, and Re).\cite{Kuc2011}
Fig.~\ref{fig:4} shows the band structures of MoS$_2$ and WS$_2$ bulk and monolayers.
When a bulk TMS$_2$ is thinned to a monolayer, there is a transition from an indirect ($\Delta$ occurs between k=$\Gamma$ and k=1/2($\Gamma$--K)) to a direct gap semiconductor ($\Delta$ occurs at k=K).
This suggests that the quantum confinement plays the key role in tailoring the electronic structure of such systems.

\begin{figure}[h!]
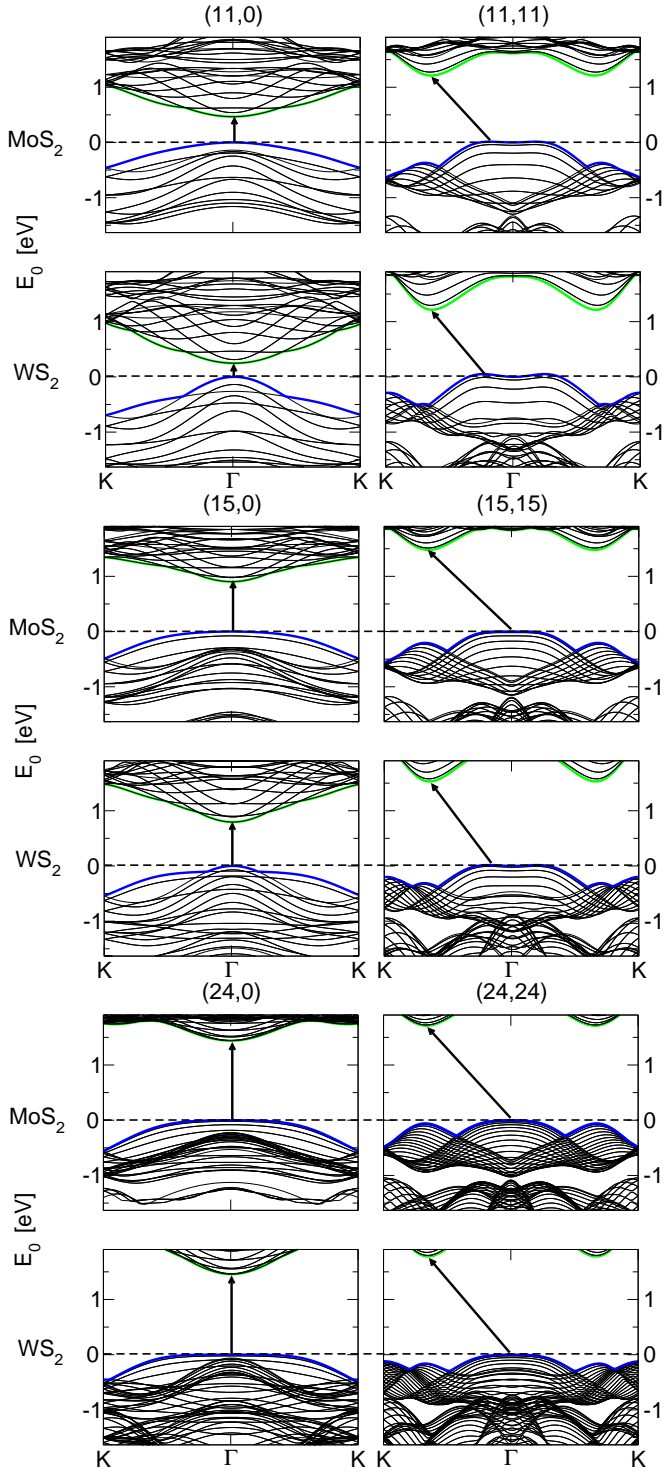

\begin{center}
\includegraphics[scale=0.33,clip]{BS_TMS2_NTs_11_eV.eps}
\includegraphics[scale=0.33,clip]{BS_TMS2_NTs_15_eV.eps}
\includegraphics[scale=0.33,clip]{BS_TMS2_NTs_24_eV.eps}
\caption{\label{fig:5}(Color online) Band structures of tubular TMS$_2$ (TM = Mo and W) with armchair and zigzag arrangement and chiral index n = 11, 15 and 24. The arrows indicate the fundamental band gap (direct or indirect) for a given system. The top of valence band (blue) and bottom of conduction band (green) are highlighted.}
\end{center}
\end{figure}

Fig.~\ref{fig:5} shows exemplary band structures around the Fermi level of TMS$_2$-NTs in zigzag and armchair configurations and chiral index n = 11, 15, and 24.
Band structures of the zigzag TMS$_2$-NTs resemble that of the corresponding single-walled carbon nanotubes (SWCNTs).
However, in contrast to SWCNTs, all the TMS$_2$-NTs are semiconducting.
While the band structures of (n,0) TMS$_2$-NTs can be related to that of a monolayer, with direct $\Delta$, the band structure of (n,n) TMS$_2$-NTs are more related to the bulk, especially for larger tube diameters.

\begin{figure}[h!]
\begin{center}
\includegraphics[scale=0.35,clip]{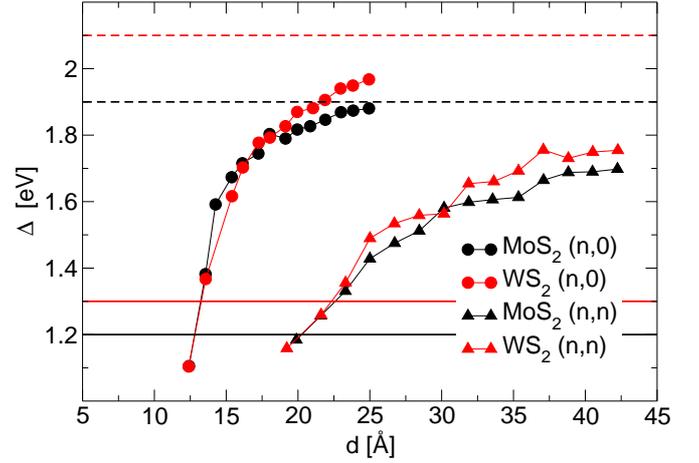}
\caption{\label{fig:6}(Color online) Calculated gap energies ($\Delta$) of (n,0) and (n,n) TMS$_2$-NTs (TM = Mo and W) with respect to the tube diameter ($d$). The corresponding layered structures are given as horizontal lines: solid--bulk, dashed--monolayer.}
\end{center}
\end{figure}

For a given tube diameter, the band gap values of zigzag NTs are larger than those of armchair NTs (see Fig.~\ref{fig:6}).
These values, in contrast to SWCNTs, increase with increasing tube diameter, going from values close to that of bulk systems (for small d) and approaching those of monolayers (for large tube sizes).

\begin{figure}[ht!]
\begin{center}
\includegraphics[scale=0.35,clip]{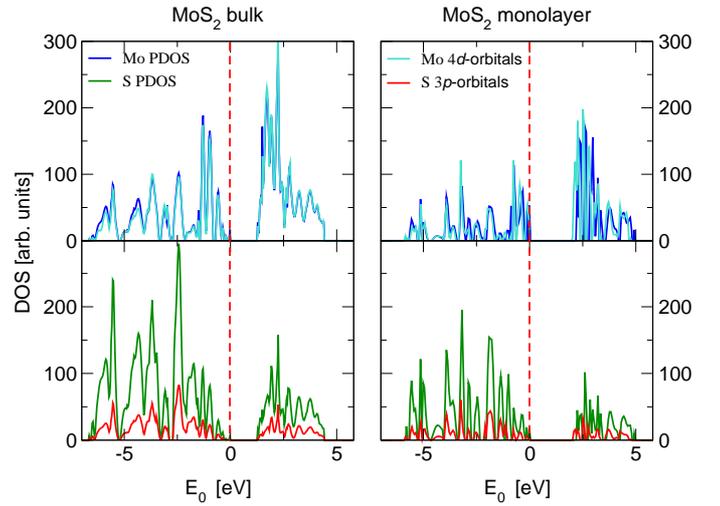}
\caption{\label{fig:7}(Color online) Partial density of states of bulk and monolayered MoS$_2$. The projections of Mo and S atoms are given together with the contributions from 4$d$ and 3$p$ orbitals of Mo and S, respectively. The vertical dashed lines indicate the the top of valence band.}
\end{center}
\end{figure}

The electronic structures of MoS$_2$ and WS$_2$ and the resulting optical properties come from the $d$-electron orbitals that dominate the valence and conduction bands (see Fig.~\ref{fig:7} for MoS$_2$).
The projected DOS (PDOS) of TMS$_2$ shows that $p$-states of sulfur atoms hybridize with the $d$-states of the transition metal atoms at the top of valence band and the bottom of conduction band.
The core states are dominated by the $s$-orbitals of the chalcogenide atom.
The PDOS of tubular structures do not change significantly from that of layered forms.


\section{Conclusions}
\label{Sec:Conclusions}

We have studied electronic properties of TMS$_2$ (TM = Mo and W) in the layered and tubular forms.
For the layered structures we have found the crossover from an indirect- to a direct-band gap semiconductor, when going from bulk to a monolayer limit.
The strain energies, originating from rolling up a monolayer to a tube, are one order of magnitude larger than in case of SWCNTs, but decrease for larger tube diameters.
These strain energies scale as $1/d^2$ ($d$ -- tube diameter), confirming also the results of Seifert et al.\cite{Seifert2000} obtained at the DFTB level.

There are no size dependent electronic irregularities of the studied TMDC-NTs, as it is found for SWCNTs.
The band gap values increase with increasing the tube size and range from that of a bulk structure to that of a monolayer.
Zigzag NTs have larger band gaps than armchair ones for a given $d$.
The very interesting phenomenon was found for the electronic band structures: zigzag NTs resemble the band structure of monolayers, while the armchair NTs (especially for large tube sizes) resemble the electronic structure of bulk systems.

\section{Acknowledgments}
\label{Sec:Acknowledgments}

AK thanks DFG -- Deutsche Forschungsgemeinschaft for financial support.

\clearpage
\providecommand*\mcitethebibliography{\thebibliography}
\csname @ifundefined\endcsname{endmcitethebibliography}
  {\let\endmcitethebibliography\endthebibliography}{}

\end{document}